\documentclass[aps,prd,twocolumn,superscriptaddress,nofootinbib,floatfix,amsmath,amssymb]{revtex4-2}

\usepackage{graphicx}
\usepackage{bm}
\usepackage{xurl}
\usepackage[colorlinks=true,linkcolor=blue,citecolor=blue,urlcolor=blue]{hyperref}
\renewcommand{\vr}{\mathbf{r}}
\usepackage{array}[=2016-10-06]

\begin{document}

\title{
Geometry-conditioned differentiable surrogate\\ with adjoint Monte Carlo sampling for scintillator detector co-design
}

\author{Lorenzo Pepa}
\email{pepalorenzo.work@gmail.com}
\affiliation{Hewlett Packard Labs, HPE Belgium, B-1831 Diegem, Belgium}
\author{Mohammad Ali Marashli}
\affiliation{Lule\aa{} University of Technology, Lule\aa{}, Sweden}
\author{Nicolás Babío Elguero}
\affiliation{Instituto de Física de Cantabria (IFCA), Universidad de Cantabria-CSIC, Santander, Spain}
\author{Tommaso Dorigo}
\affiliation{Istituto Nazionale di Fisica Nucleare -- Sezione di Padova, via F. Marzolo 8, 35131 Padova, Italy}
\affiliation{Lule\aa{} University of Technology, Lule\aa{}, Sweden}
\author{Irene Fagnani}
\affiliation{Lule\aa{} University of Technology, Lule\aa{}, Sweden}
\author{Carlo Gulizia}
\affiliation{Departamento de Física and ICTEA, Universidad de Oviedo, Oviedo, Spain}
\author{Pablo Martinez Ruiz del Arbol}
\affiliation{Instituto de Física de Cantabria (IFCA), Universidad de Cantabria-CSIC, Santander, Spain}
\author{Fredrik Sandin}
\affiliation{Lule\aa{} University of Technology, Lule\aa{}, Sweden}
\author{Pietro Vischia}
\affiliation{Departamento de Física and ICTEA, Universidad de Oviedo, Oviedo, Spain}

\date{\today}

\begin{abstract}
%
Optimizing light-guide geometries for scintillator-based imaging detectors requires repeated photon-transport evaluations, making direct Monte Carlo simulation computationally costly.
%
Differentiable surrogates can accelerate these evaluations and support detector–readout co-design aimed at improving information extraction from particle interactions.
In this work we describe a surrogate model for photon transport. We develop it starting from a modular, event-driven Monte Carlo model for the generation and transport of scintillation photons, and a reciprocal Monte Carlo construction (an adjoint model) of a spatially and temporally resolved optical response field. 
The latter evaluates the collectible light output and its temporal structure from a scintillator volume by launching weighted photon trajectories from the light-guide readout face. We apply this approach to a scintillator cell with two mirrored, curved, single-clad light guides. We employ the adjoint model to train a geometry-conditioned neural surrogate that predicts collection probabilities and conditional transport-time distributions across a constrained guide-design family. The surrogate is natively differentiable, lending itself to be an element of a differentiable pipeline for gradient-based optimization.
A forward-reciprocal benchmark and validation on held-out geometries establish the accuracy of the trained model and identify its limitations.
The developed tools provide a foundation for rapid optical-response studies and detector-readout co-design.
\end{abstract}

\maketitle

\section{Introduction}
\label{sec:intro}

Scintillation detectors convert the energy deposited in their volume by ionizing particles into optical photons. In typical particle detectors, the observable quantities one tries to estimate from the process include the number of collected photons at the sensor end 
and their arrival time; wavelength and polarization are additional observable quantities that may be needed to better characterize the system performance in specific applications. The number of photons is then subjected to a calibration procedure that yields an estimate of the total energy deposited in the scintillator by ionizing particles, assuming the transport efficiency to the detection sensors is known. Additional information about the time structure as well as about the spatial topology of the ionization deposits can in principle be extracted from the arrival time  distribution of the detected photons if the transport time and excitation-de-excitation processes are precisely accounted for. 

The inference of the map of ionization energy distribution in the scintillation volume as a function of time, given the observable features mentioned above, constitutes a complex inverse problem. Its solution is important in modern calorimeter applications because the topology of energy depositions may give access to detailed properties of particle showers, and through those enable inference on the identity of the originating particles~\cite{devita1,devita2}.
In realistic setups, the problem requires the detailed accounting of all possible propagation trajectories of the detected photons inside light guides and scintillator volume, as well as all involved phenomena, including absorption, scattering, or reflection at boundaries. There is, in summary, much more to account for than a single position-independent light-collection coefficient.

The \textsc{Geant4} toolkit is widely used to model particle interactions and energy deposition in detector materials~\citep{Agostinelli2003,Allison2006,Allison2016}. It also provides a detailed optical-photon physics package, including scintillation and Cherenkov photon production, absorption, scattering, wavelength shifting, and interactions at optical boundaries~\citep{Geant4Optical}. Because energetic interactions produce many photons,  tracking the latter and their numerous surface interactions may dominate computing time for particular detector geometries. Consequently, an application may elect to stop detailed particle transport at the energy-deposit stage and handle optical response through precomputed maps, parameterizations, or separate specialized simulation. The development of GPU optical-photon tools is another indication of the computational importance of this problem~\citep{Blyth2017}.

Repeated simulation becomes even more demanding when the optical configuration is itself an optimization variable. Changing scintillator dimensions, guide width, curvature or path, optical interfaces, or readout location can alter the relationship between energy deposition and detectable signal. In a program of optimization of detector design such as those considered in Ref.~\cite{Dorigo2023}, one wishes to evaluate a large number of such choices under a common model, and ideally to obtain an optical response that varies smoothly with design parameters. Differentiable rendering and inverse light-transport methods provide valuable examples of this broader trend~\citep{NimierDavid2019}. Nevertheless, a model that can be differentiated does not necessarily predict the derivatives of the physical response correctly. Differentiating the transport itself is complicated by sharp boundaries, total-internal-reflection thresholds, and changes of ray-path topology; a smooth learned model need not follow the true response on small scales. We test this for our surrogate against finite differences of the adjoint model (Sec.~\ref{sec:res_derivatives}).
The issues outlined above arise concretely in the PHINDER program~\cite{phinder}. The project considers scintillator cells coupled to light guides and optoelectronic readout, and aims to perform co-design of light transport and neuromorphic signal processing. The goal is to optimally yield topological information on energy deposition processes needed to infer more properties on the particle showers. The present work, developed within PHINDER, addresses the optical modeling problem of how to preserve the source-position-dependent collection and timing response of curved light guides while making repeated evaluations sufficiently economical for design studies.

\begin{figure*}[t]
  \centering
  \includegraphics[width=0.95\textwidth]
  {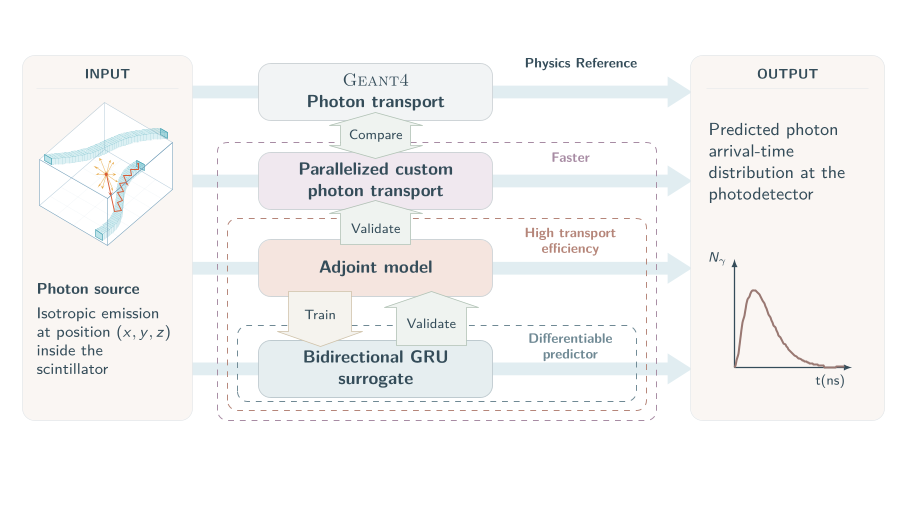}
  \caption{Overview. Photon transport from an emission point to the readout is computed with \textsc{Geant4}, with a dedicated
  forward engine, and with the adjoint model, which gives the response for all emission points at once. The adjoint model
  is validated against \textsc{Geant4} and the forward engine and generates the training data of a differentiable neural surrogate.}
  \label{fig:overview}
\end{figure*}

We develop this approach in three stages, illustrated in Fig.~\ref{fig:overview}. First, a dedicated event-driven Monte Carlo simulator (Sec.~\ref{sec:forward}) provides the physically interpretable \emph{forward} optical reference; its agreement with Geant4 confirms the physics that the other two stages share. Second, an adjoint model (Sec.~\ref{sec:adjoint}) uses reciprocity to transport weighted rays backward from the
readout face. Every traced path scores all the voxels it crosses, correctly accounting for propagation time and transport probability; at equal accuracy on the information reaching the readout, the adjoint model proves to be three orders of magnitude faster to compute than the
simulation of emitted photons of the forward model. Third, a geometry-conditioned neural surrogate (Sec.~\ref{sec:surrogate}), trained on an
atlas of adjoint fields, predicts the response of a new design in milliseconds and is differentiable with respect to the
design parameters, so that hundreds of geometries can be screened per second, and the main design dimensions effectively optimized by gradient descent. Section~\ref{sec:results} validates the adjoint model and the surrogate against \textsc{Geant4}, quantifies the accuracy
of the surrogate and of its derivatives, and compares the cost of all methods on matched tasks.

\section{Related approaches}
\label{sec:related}

\paragraph*{Forward optical-photon simulation.}

While \textsc{Geant4} can track optical photons in detail if the user configures it to do so~\cite{Geant4Optical}, this step is often executed by custom-made, geometry-specific programs, which may be easier to inspect and modify. The PHOTON code~\cite{Tickner2007}, for example, was written
for scintillation-detector optics and includes configurable boundaries, complex refractive indices and thin films. Because the transport of optical photons is an easy-to-parallelize task, this has become the method of choice when GPUs are available; e.g., Opticks couples \textsc{Geant4} to GPU photon generation and propagation with hardware ray tracing~\cite{Blyth2017}. 
GPU acceleration reduces the computational cost of forward optical transport by propagating many photons in parallel, but still requires each photon to be tracked individually. An alternative is to precompute the detector's optical response as a function of emission position, allowing the same response field to be reused for different events without repeating photon transport~\cite{GarciaGamez2021}.

\paragraph*{Reverse and adjoint transport.}

Adjoint formulations are well established in neutron transport~\cite{Spanier1969}, radiative transfer, optical imaging, and
computer graphics: when a detector subtends a small region of phase space, most photons launched from possible emission
points miss it, making forward transport a very inefficient way to test acceptance and determine propagation properties. The adjoint view instead considers which emission points and directions can contribute to the measurement:
trajectories launched from the detector end sample an importance field that can be combined with any source distribution. For example, a forward-adjoint fluorescence model using this concept was developed and tested experimentally by Crilly \emph{et al.}~\cite{Crilly1997}, and
coupled forward and adjoint fields were used to compute optical sensitivity in scattering media~\cite{Gardner2014}. In
graphics, Veach and Guibas~\cite{Veach1995} showed that a light path can be sampled from the light or from the camera with very different variance. In addition, inverse renderers such as Mitsuba~2~\cite{NimierDavid2019} differentiate light transport with respect to scene parameters.

\paragraph*{Learned surrogates.}

Machine learning models have become an important alternative to computationally expensive Monte Carlo detector simulation for particle physics applications. Early applications, such as CaloGAN~\cite{Paganini2018}, demonstrated that generative neural networks could reproduce aspects of calorimeter shower development at a fraction of the cost of detailed simulation. More recent approaches employ normalizing flows, diffusion models and flow matching to improve the fidelity of simulated detector responses. The CaloChallenge benchmark ~\cite{Krause2025calochallenge} has provided a systematic comparison of these methods, highlighting both their progress and the importance of evaluating accuracy using physically relevant observables rather than aggregate statistical measures alone. Besides accelerating event simulation, differentiable generative surrogates have become the enabling tool to allow for gradient-based detector-design optimization~\cite{Schmidt2025aido,Nguyen2026}.

A related but distinct approach is to learn the detector response as a continuous function of the source coordinates or other physical parameters. Neural operators, including DeepONet~\cite{Lu2021}, provide a general framework for approximating mappings between functions and have attracted considerable interest in scientific simulation and inverse design~\cite{Azizzadenesheli2024}. For optical detectors, neural representations can replace or compress conventional photon-visibility lookup tables. For example, sinusoidal representation networks have been investigated for modeling scintillation-photon propagation in liquid-argon time projection chambers~\cite{Gasiorowski_2024}. More recently, a differentiable neural representation of photon propagation in the IceCube neutrino telescope has been developed to predict both photon yields and arrival-time distributions, allowing the optical response of complex events to be constructed by superposition~\cite{Yu2026candela}. Generative approaches have also been applied directly to optical photon transport in scintillator arrays~\cite{Naunheim2026optigan}.

These developments motivate the use of learned optical response functions as an alternative to repeatedly propagating
individual photons. Existing optical surrogates, however, represent one fixed detector, such as a liquid-argon time
projection chamber~\cite{Gasiorowski_2024} or a neutrino telescope~\cite{Yu2026candela}, and must be retrained when its geometry changes.
Surrogates that take the design itself as input are used for inverse design in nanophotonics, including nanophotonic
scintillators~\cite{Regev2026}, and for shape optimization~\cite{Sun2026gano}, but, to our knowledge, not for the
light-collection and timing response of scintillator light guides. Likewise, adjoint Monte Carlo is standard in neutron
and gamma transport, and \textsc{Geant4} provides it for gamma rays, electrons and ions~\cite{DESORGHER2010247}, but optical
response maps of scintillators are usually obtained by forward simulation or by parameterizations fitted to
it~\cite{GarciaGamez2021}. The approach pursued here combines the two. It learns the one-photon response operator,
separating collection probability from the conditional arrival-time distribution; the light-guide geometry is an input
to the network; and its training data are produced by reciprocal Monte Carlo, which gives the response of every emission
point of a design in one run. The aim is a compact representation of the optical response across a continuous family of
detector designs, suitable both for fast event simulation and for studying the sensitivity of the response to changes in
geometry.

\section{Forward Monte Carlo transport}
\label{sec:forward}

\subsection{Photon production and transport}

The forward simulator separates the generation of optical photons from their propagation and collection. Its input is a
list of energy deposits with their positions and times, taken from a \textsc{Geant4} simulation of incident particles or chosen by the
user to probe specific paths. For a deposit $\Delta E$ and scintillation yield $Y$ the number of photons is drawn as $N_\gamma \sim \mathrm{Poisson}(Y\,\Delta E)$. The photons are emitted isotropically;
their emission points can be spread along the track segment of the deposit, and each can be delayed by a time drawn
from the scintillation profile. For the response fields of Sec.~\ref{sec:adjoint} the delay is
set to zero, so that the fields describe optical transport time only; the emission-time profile is applied afterwards by convolution, Eq.~(\ref{eq:superposition}).

The simulation of a photon carries forward its position, direction, elapsed time, current material and status. At each step the nearest boundary along the ray is found and the photon moves there unless it is absorbed first; a path length $\ell$ in a medium of index
$n$ adds $\Delta t = n\ell/c$ to the elapsed time. We use geometrical optics with a constant refractive index per material, so dispersion and wavelength-dependent emission and absorption are not included in the present version of the software. Because photons jump from boundary to boundary, the computation cost
is set by the number of surface interactions, which is large for photons trapped by total internal reflection.

For an absorption coefficient $\mu$ the absorption distance is drawn as $\ell_{\rm abs} = -\ln\xi/\mu$, with a uniformly sampled $\xi \sim U(0,1)$;
re-emission is not modeled. At an interface the transmitted direction follows Snell's law,
$n_1\sin\theta_i = n_2\sin\theta_t$, and the reflection probability is the average of the two Fresnel reflectances,
\begin{align}
  R_s &= \left(\frac{n_1\cos\theta_i - n_2\cos\theta_t}{n_1\cos\theta_i + n_2\cos\theta_t}\right)^{2}, \nonumber\\
  R_p &= \left(\frac{n_1\cos\theta_t - n_2\cos\theta_i}{n_1\cos\theta_t + n_2\cos\theta_i}\right)^{2}, \nonumber\\
  R   &= \tfrac12\,(R_s + R_p), \qquad T = 1 - R .
  \label{eq:fresnel_avg}
\end{align}
A Bernoulli draw selects reflection or transmission; total internal reflection occurs when Snell's law has no solution.
Averaging over polarization is an approximation whose effect we measure in Sec.~\ref{sec:res_geant4}. A photon is followed
until it is detected, absorbed, or escapes.

\subsection{Detector geometry and readout}
\label{sec:geometry}

\begin{figure}
    \centering
    \includegraphics[width=1\linewidth]{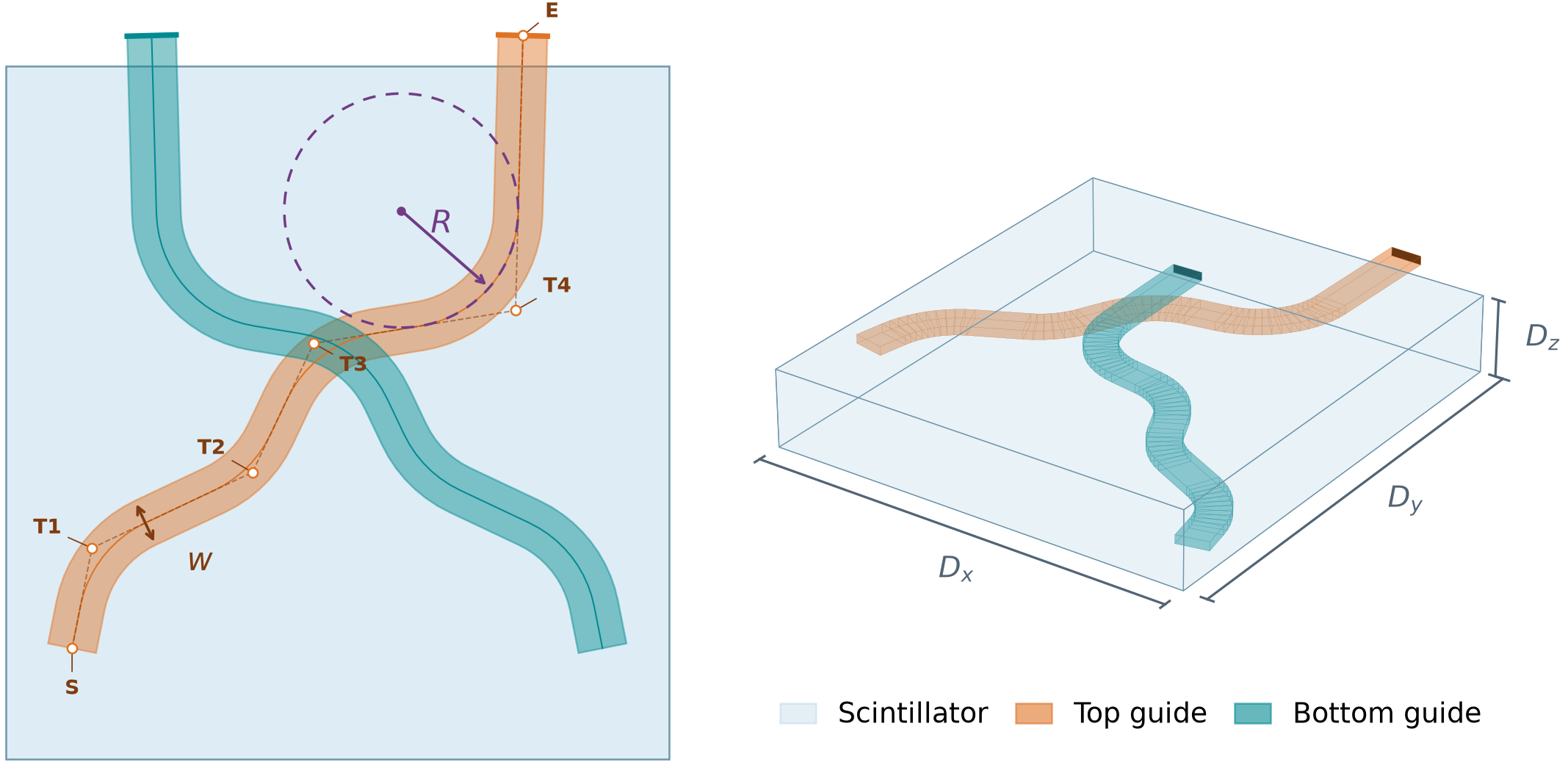}
    \caption{An EJ-204 cell of the design family with its two mirrored curved guides (core and cladding) and their readout
  ends. Left: top view of the mirrored light guides, showing the start (S), exit (E), internal control vertices (T1–T4), guide width (w), and common bend radius (R). The dashed circle illustrates the centerline curvature of a bend. Right: three-dimensional view of the scintillator and guides, with scintillator dimensions $D_x$, $D_y$, and $D_z$. Orange and teal denote the top and bottom guides, respectively; dark end faces indicate the readouts.}
    \label{fig:cell}
\end{figure}

The detector is an EJ-204 plastic scintillator cell of dimensions $D_x\times D_y\times D_z$ read out by two light guides on
its large faces (Fig.~\ref{fig:cell}). Guide~0 lies on the $+z$ face; guide~1 is its mirror image under $x \to D_x - x$ on
the $-z$ face, so that the response of channel~1 follows exactly from that of channel~0. Each guide has a rectangular
cross-section of width $w$ and thickness 0.10~cm, with a core and a 10~$\mu$m cladding on all sides. Its path is a polyline
through six vertices (start, four turning points, terminal) whose corners are replaced by circular fillets of radius $R$,
built from 16 straight facets per $90^\circ$. The guide extends 0.25~cm beyond the cell into a terminal core segment;
photons reaching its end face, the readout port, are counted as detected with unit efficiency. Everything else is air;
any photon transmitted into air is terminated (Sec.~\ref{sec:res_geant4} measures the effect of this choice).
Table~\ref{tab:materials} lists the optical constants.

The design family we consider in this work has 13 continuous parameters: the cell dimensions ($D_x, D_y \in [4,6]$~cm, $D_z \in [0.7,1.3]$~cm),
the guide width ($w \in [0.4,0.8]$~cm), the fillet radius through $R/w \in [1.5,2.5]$, and the $x$ and $y$ coordinates of the
four turning points (12-88\% of $D_x$ and 22-88\% of $D_y$, with ordered $y$). These parameters do not span the totality of imaginable configurations, yet they provide a useful baseline for the development and test of a full co-design pipeline.

\begin{table}[t]
\centering
\small
\caption{Optical constants used by all models. Indices are wavelength independent; photons have 3.04~eV (408~nm).}
\label{tab:materials}
\begin{tabular*}{\columnwidth}{@{\extracolsep{\fill}}>{\raggedright\arraybackslash}p{0.35\columnwidth}cc}
\hline
Material & Index & Absorption length [cm] \\
\hline
EJ-204 scintillator & 1.58 & 160 \\
Guide core          & 1.59 & 200 \\
Guide cladding      & 1.49 & 500 \\
Air                 & 1.00 & N/A \\
\hline
\end{tabular*}
\end{table}

The engine is written in JAX~\cite{Bradbury2018} and vectorizes over photons. It serves here as an independent reference
implementation of the optical model; Appendix~\ref{app:figures} compares its run time with that of \textsc{Geant4} on CPU and GPU.
Dedicated GPU tools with hardware ray tracing, such as Opticks~\cite{Blyth2017}, can be faster still.

\section{Adjoint response fields}
\label{sec:adjoint}

\subsection{The one-photon optical Green function}

We seek the response to one photon emitted isotropically at position $\vr$ in a geometry $G$. For readout channel $c$,
\begin{equation}
  h_c(t\,|\,G,\vr) = p_c(G,\vr)\, f_c(t\,|\,G,\vr)\,,
\end{equation}
where $p_c$ is the detection probability and $f_c$ the arrival-time density given detection, normalized to unity; $h_c$ is
then the density of detected arrivals per emitted photon. For an event with energy deposits at positions $\vr_k$ and times $t_k$
producing on average $\mu_k$ photons, independent photon transport gives the expected detected intensity
\begin{equation}
  \lambda_c(t) = \sum_k \mu_k \left[s_k * h_c(\cdot\,|\,G,\vr_k)\right](t - t_k)\,,
  \label{eq:superposition}
\end{equation}
where $s_k$ is the normalized emission-time profile and $*$ denotes convolution. The computing cost of Eq.~(\ref{eq:superposition})
depends on the number of deposits rather than on the number of photons, thus making a response field useful for event
simulation (Sec.~\ref{sec:res_cost}).

\subsection{Reverse launch and normalization}

\begin{figure*}[t]
  \centering
  \includegraphics[width=\textwidth]{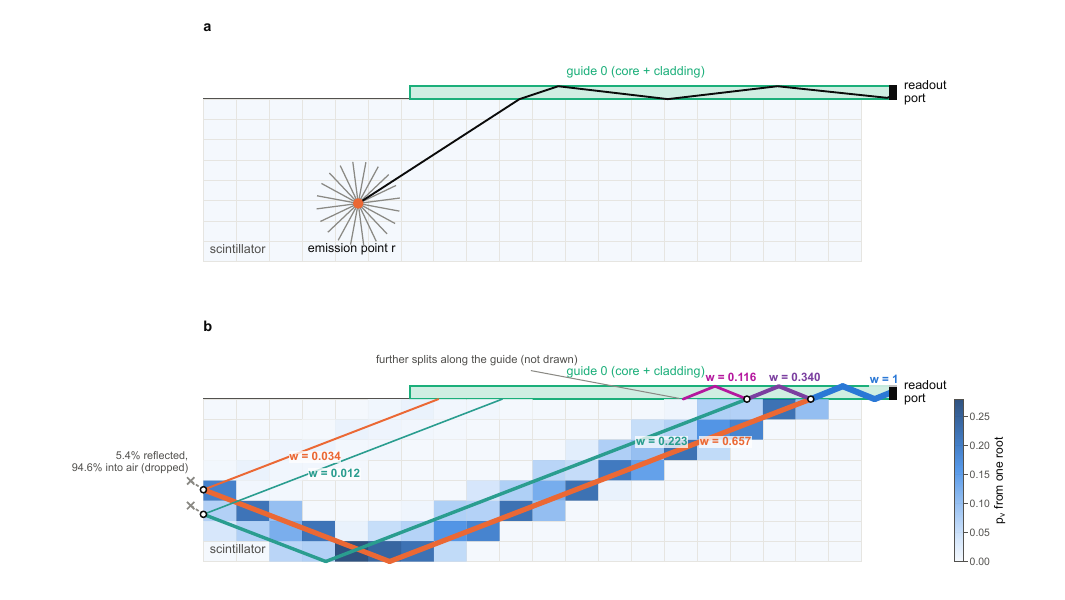}
    \caption{Forward and reverse sampling. (a) A forward simulation emits photons from one point; about 1\% reach the port.
  (b) One reverse root traced through a side view of a test cell  ($5.1\times1.26$~cm, voxels of
  0.0085~cm$^3$), in two dimensions and without the second guide. Blue: the root in the guide core ($w=1$ at the port); orange: the branch entering the scintillator;
  purple: the branch staying in the guide; teal: its second entry into the cell. Line width shows the branch weight;
  $\times$ marks branches that leave into air and are dropped. Color bar: the estimate of $p_v$ from this root alone.}
  \label{fig:reciprocal}
\end{figure*}

For a given channel, rays (``roots'') are launched from the readout face back into the guide, uniformly over the port area
$A_{\rm port}$ and with the cosine-weighted direction measure $q(\mathbf x, \Omega) = \cos\theta/(\pi A_{\rm port})$.
The four numbers that set the start point on the port and the direction of each root ($\cos\theta = \sqrt{u_2}$,
$\phi = 2\pi u_3$) are taken from a scrambled Halton sequence~\cite{Halton1960}, which covers the port and the
directions more evenly than independent random numbers and so reduces the noise for a given number of roots. The reverse field is converted to the forward detection probability by
the phase-space factor
\begin{equation}
  p_c(\vr) = \frac{A_{\rm port}}{4}\left(\frac{n_{\rm port}}{n_{\rm scint}}\right)^2 \phi^\dagger_c(\vr),
\end{equation}
which follows from the invariance of the \'etendue $n^2\cos\theta\,dA\,d\Omega$. Here $\phi^\dagger_c$ is the weighted
reverse track length per unit voxel volume and per root, $\hat\phi^\dagger_{c,j} = (N V_j)^{-1}\sum_{i}\sum_{s} w_{is}\ell_{is}$,
where $\ell_{is}$ is the length of segment $s$ of root $i$ inside voxel $j$, computed from exact voxel-plane crossings, and
$w_{is}$ its weight.  Binning the same contributions by total reverse travel time gives $h_c$. Each root scores every voxel that is crossed by one of its branches,(Fig.~\ref{fig:reciprocal}b), so that a single run gives the response for all possible emission points at once. Figure~\ref{fig:reciprocal}b follows one root through a side view of a test cell. It starts at the port with weight
$w=1$ and travels back through the guide core. Where the guide lies on the scintillator, the core-cladding-scintillator
interface splits it (Sec.~\ref{sec:branching}): 0.66 of the weight enters the scintillator and 0.34 stays in the guide,
which splits again at its next contact (0.22 into the cell, 0.12 onward). The scintillator branch meets the bottom face beyond the $39^\circ$ critical angle of the scintillator-air interface, so the reflection is total and
nothing is lost. Then the branch meets the side wall at $21^\circ$, where 5.4\% is reflected and 94.6\% leaves into air and is
dropped. The weak reflected branches ($w = 0.034$ and 0.012) reach the top face, where 82\% of each enters the guide.
Between interfaces the weight falls slowly through bulk absorption. Each scintillator voxel crossed by a branch receives
$w\ell/V$ in the bin of its travel time, one term of $\hat\phi^\dagger_{c,j}$; multiplied by the phase-space factor
($0.0105$~cm$^2$ here) this is the one-root estimate of the detection probability shown in color in the figure. The map is the mean
of these estimates over all $N$ roots, most of which never reach a given voxel.

\begin{figure*}[t]
  \centering
  \includegraphics[width=\textwidth]{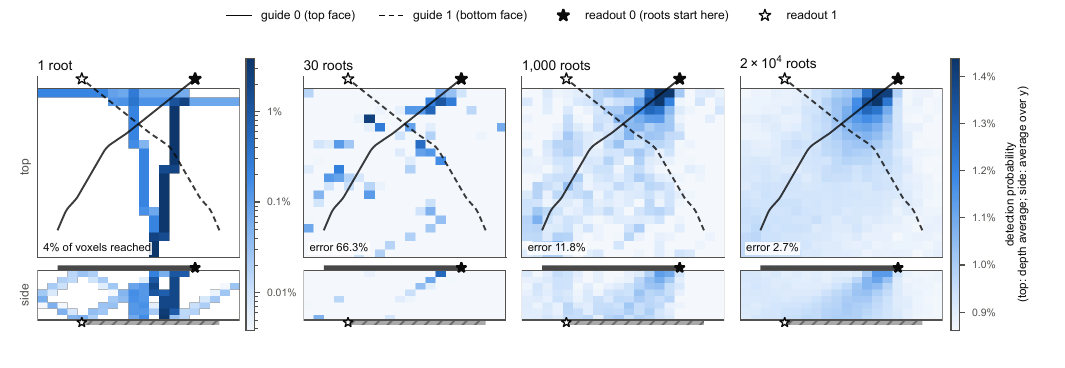}
  \caption{Detection probability of channel~0 in a test geometry ($5.1\times4.2\times1.26$~cm$^3$), estimated from the first 1, 30,
  1000, and $2\times10^4$ roots, seen from the top (averaged over depth) and from the side (averaged over $y$), with both
  guides drawn: guide~0 on the top face, guide~1 (hatched in the side view) on the bottom face.
  The single root has its own color scale, so that its weaker branches are visible; the other
  panels share the same color scale. Error: $p$-weighted RMS of the relative difference to a $2\times10^5$-root map, over the voxels
  reached.}
  \label{fig:backtrace_maps}
\end{figure*}
\subsection{Branching, absorption and limits}
\label{sec:branching}

At an interface a reverse ray can continue as both a reflected and a transmitted branch, with weights $wR$ and $wT$, or
choose one with probabilities $R$ and $T$ (analog transport). We split when both children would carry at least a fraction
$\tau = 0.005$ of the parent weight and use the analog choice otherwise; both procedures are unbiased. One root is therefore a tree of branches. Figure~\ref{fig:backtrace_maps} shows this in a test geometry, together with
how the map builds up as roots accumulate: the first root scores 33 straight segments in the scintillator, with weights
from 0.81 down to 0.006, bouncing between the two faces of the cell, and reaches 4\% of the voxels; after
$2\times10^4$ roots the map covers the whole cell.
Bulk absorption is a
continuous weight $e^{-\mu\ell}$. Random numbers are computed from the index of the root, branch and step rather than drawn in sequence, so that two
slightly different geometries can be traced with the same random numbers. Their rays then follow the same paths until
the change in geometry makes them differ, and much of the noise cancels in the difference of the collected light; for
arrival times the paths separate too quickly for this to help much (Sec.~\ref{sec:res_derivatives}). A branch is followed until it escapes into air, leaves
through the readout port, or reaches a maximum set to 5000 boundary interactions or a propagation time of 80~ns. 

\subsection{The geometry atlas}
\label{sec:atlas}

Designs are drawn from a 13-dimensional Sobol sequence~\cite{Sobol1967} (one dimension per free parameter of the geometry), and are checked before transport: fillets must fit on
their segments, consecutive segments must not fold back, and the guide must stay on the cell face. Only 2.0\% of 139,012
candidates pass the selection; the valid designs are thinned by greedy maximin selection into three banks of 84, 250 and 350 geometries.
For each geometry we compute the channel-0 field on a $20\times20\times8$ voxel grid with 240 arrival-time bins of 0.25~ns,
using $N = 2\times10^4$ roots (3.6~s per geometry on 14 cores; voxel error 2.4\%, Sec.~\ref{sec:res_adjoint}). Forty test and 16 validation
geometries are set aside, and 40 further test geometries are drawn from an independent Sobol sequence. The 80 test geometries are traced with $2\times10^5$ roots and
independent seeds to serve as references.

\section{Geometry-conditioned surrogate}
\label{sec:surrogate}
\begin{figure}[t]
  \centering
  \includegraphics[width=\columnwidth]{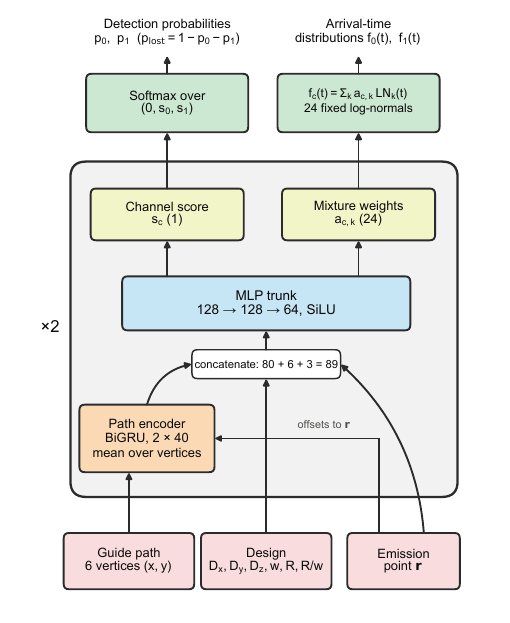}
  \caption{Surrogate network. The shaded block is evaluated twice with the same weights: at the emission point $\vr$ for
  channel~0 and at its mirror image $\vr'$ for channel~1. A softmax over $(0, s_0, s_1)$ gives the detection
  probabilities, and each channel's arrival-time distribution is a mixture of 24 fixed log-normal components. Layer sizes
  in Table~\ref{tab:arch}.}
  \label{fig:arch}
\end{figure}

\begin{table}[t]
\centering
\small
\caption{Surrogate layers for one evaluation; channel~1 reuses the same weights at the mirrored point.}
\label{tab:arch}
\begin{tabular}{>{\raggedright\arraybackslash}p{0.35\columnwidth}ccr}
\hline
Layer & Size & Activation & Parameters \\
\hline
Guide vertices & $6\times5$ & -- & -- \\\noalign{\vskip 3pt}
BiGRU, mean over vertices & 80 & -- & 11\,280 \\\noalign{\vskip 3pt}
Concatenate design, $\vr$ & 89 & -- & -- \\\noalign{\vskip 3pt}
Dense & 128 & SiLU & 11\,520 \\\noalign{\vskip 3pt}
Dense & 128 & SiLU & 16\,512 \\\noalign{\vskip 3pt}
Dense & 64 & SiLU & 8\,256 \\\noalign{\vskip 3pt}
Channel score $s_c$ & 1 & -- & 65 \\\noalign{\vskip 3pt}
Mixture logits & 24 & softmax & 1\,560 \\\noalign{\vskip 3pt}
Fate (loss, ch.~0, ch.~1) & 3 & softmax & -- \\\noalign{\vskip 3pt}
Timing $f_c(t)$ & 240 bins & log-normal mix & -- \\
\hline
Total & & & 49\,193 \\
\hline
\end{tabular}
\end{table}

The network maps a geometry $G$ and a normalized emission point $\vr = (x/D_x, y/D_y, z/D_z)$ to probabilities and arrival times of photons at the light guides end,
$\{p_0, f_0(t), p_1, f_1(t)\}$ (Fig.~\ref{fig:arch}, Table~\ref{tab:arch}). The six guide vertices, with their
displacements from the emission point, are encoded by a bidirectional gated recurrent unit~\cite{Cho2014} (40 units per
direction, averaged over vertices); the code is concatenated with the design scalars $(D_x, D_y, D_z, w, R, R/w)$ and $\vr$
and passed through a perceptron with layers of 128, 128 and 64 SiLU units. Only two of $w$, $R$ and $R/w$ are independent
(the atlas samples $w$ and $R/w$, Sec.~\ref{sec:geometry}); the third is passed as a derived feature. 
Two heads give a channel score and 24 mixture weights. Channel~0 is evaluated at the emission point, giving $s_0$ and
$f_0$; channel~1 is the same network, with the same weights, evaluated at the mirror image $(D_x - x, y, D_z - z)$ of
the emission point, giving $s_1$ and $f_1$, which makes the mirror symmetry exact. With a fixed score of 0 for loss, a
softmax over $(0, s_0, s_1)$ gives $p_c = e^{s_c}/(1 + e^{s_0} + e^{s_1})$, so that $p_0 + p_1 + p_{\rm lost} = 1$;
each $f_c$ is a mixture of 24 fixed log-normal components (medians 0.12-57.5~ns), so it is normalized by construction.
The network has 49,193
parameters. We use a recurrent encoder because it accepts guides with any number of vertices; this flexibility will be exploited when we will study other guide shapes in future extensions of this work. For the fixed six-vertex family studied here, a perceptron performs similarly (Sec.~\ref{sec:res_surrogate}). 

The loss is
\begin{equation}
\begin{split}
  \mathcal L ={}& \frac{\sum_v \rho_v \sum_c H(\ln\hat p_{c,v} - \ln p_{c,v})}{\sum_v \rho_v}\\
  &+ 0.55\,\frac{\sum_v w_v\, D_{\rm KL}(f_{0,v}\,\|\,\hat f_{0,v})}{\sum_v w_v}\\
  &+ 0.15\,\frac{\sum_v w_v\, H\!\left[(\hat{\bar t}_v - \bar t_v)/5~{\rm ns}\right]}{\sum_v w_v},
\end{split}
\label{eq:loss}
\end{equation}
summed over the voxels $v$ of a batch, where hats denote network predictions, $H$ is the Huber function with unit
threshold, $D_{\rm KL}$ the Kullback-Leibler divergence over the 240 time bins of 0.25~ns, and $\bar t_v$ the mean
arrival time. The weight $\rho_v = 1/[1+(\epsilon_v/0.35)^2]$, limited to $[0.05, 1]$, reduces the influence of voxels
with a large relative standard error $\epsilon_v$ of the atlas value $p_{0,v}$, and the timing terms use
$w_v = \rho_v \min[1, \max(0.05, p_{0,v}/0.01)]$, so that voxels with little detected light, whose arrival-time
distributions are noisy, count less.
We train with
AdamW~\cite{Loshchilov2019} (learning rate $2\times10^{-3}$ with cosine decay, batch 512) for 240 epochs and keep the
last-epoch weights; training on 628 geometries takes 23~min on the CPU.

\section{Results}
\label{sec:results}

Unless stated otherwise, all computations and timings used an Intel Core Ultra~5~245 CPU (14 cores) without a GPU. Voxel-level accuracy is measured against the adjoint field with $2\times10^5$ roots, which is
itself validated against \textsc{Geant4} below; a \textsc{Geant4} map of the same precision would take hours per geometry. The voxel error
of a map $\hat p$ is the $p$-weighted RMS of $(\hat p_v - p_v)/p_v$ over all voxels of all test geometries, with the
reference noise subtracted in quadrature; the same measure is used for the mean arrival time $\bar t$ and the
arrival-time spread $s_t$. The reference noise (0.73\% for $p$, 0.54\% for $\bar t$ and 0.36\% for $s_t$) is measured
from pairs of independent replicate maps (Sec.~\ref{sec:res_adjoint}). The layer-wise RMS values displayed in Fig. \ref{fig:maps} instead describe the observed residuals, without subtraction of reference noise in quadrature.

\subsection{Agreement with \textsc{Geant4}}
\label{sec:res_geant4}

We implemented the cell in \textsc{Geant4}~11.4.1 through its Python bindings~\cite{geant4pybind}, using the same triangulated
scene, the constants of Table~\ref{tab:materials}, and only optical absorption and boundary processes
(Appendix~\ref{app:geant4}). To match our model, the polarization is redrawn before every interface, so that reflection
occurs with probability $(R_s+R_p)/2$, and photons entering air are terminated. In eight test geometries, photons were
emitted uniformly in five regions: the whole cell ($4\times10^5$ photons) and four $2\times2$-voxel blocks
($2\times10^5$ each) next to guide~0, below it, at the center and in the far corner. The forward engine was run on four of
these geometries with four to five times fewer photons. We compare channel~0. \textsc{Geant4} simulates channel~1 independently,
and the mirrored adjoint map agrees with it as well ($\chi^2 = 29$ for 40 values), which confirms the mirror symmetry used
by the adjoint and the network.

For each region, the ratio $r_i = m_i/g_i$ of a method's captured fraction to that of \textsc{Geant4} is compared with its
uncertainty,
\begin{equation}
  \chi^2 = \sum_i \left(\frac{r_i - 1}{\sigma_i}\right)^2, \qquad
  \frac{\sigma_i}{r_i} = \sqrt{\left(\frac{\sigma_{m,i}}{m_i}\right)^2 + \left(\frac{\sigma_{g,i}}{g_i}\right)^2},
  \label{eq:chi2}
\end{equation}
If the differences are explained by the uncertainties alone, each term is about one on average, and $\chi^2$ is close
to the number of values $n$, within about $\sqrt{2n}$ ($40\pm9$ for 40 values). For \textsc{Geant4} and the
forward engine, $\sigma$ can be approximated by the Wald approximation to the binomial error, $\sqrt{q(1-q)/n}$ of the captured fraction $q$ for $n$ emitted photons
(median 1.7\% and 3.6\%); for the forward engine, the spread between batches of photons agrees with it. For the adjoint,
$\sigma$ is measured from three independent $2\times10^4$-root maps of each geometry: the differences of their region
means, pooled over the eight geometries, are scaled by $\sqrt{10}$ to the $2\times10^5$-root reference (median 0.3\%).
The network has no sampling noise, but its prediction depends on the random initialization and ordering of the training;
we take as its uncertainty the spread between three networks trained with different seeds, pooled over all 40 values
(1.8\%). This does not include errors common to all three networks, such as noise in the training maps or limited model
capacity.

\begin{figure*}[ht]
  \centering
    \includegraphics[width=\textwidth]{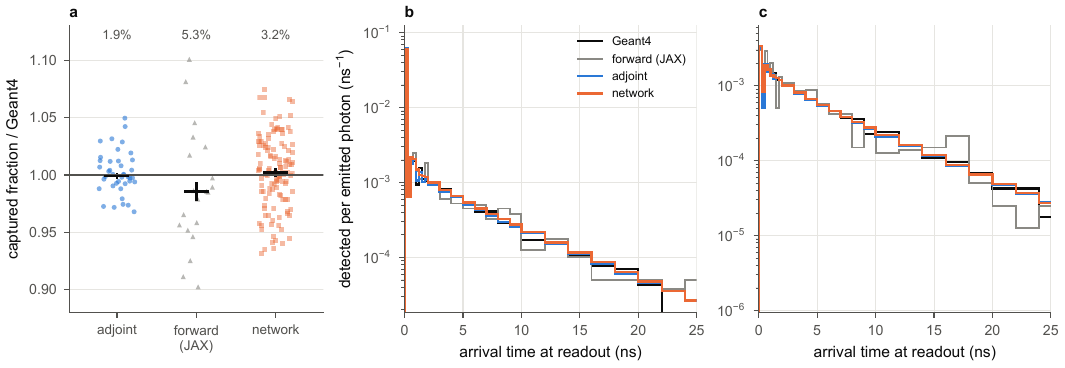}
  \caption{Comparison with \textsc{Geant4}, channel~0. (a) Captured fraction relative to \textsc{Geant4} in each emission region for the
  adjoint field, the forward engine and the network (8 test geometries; 4 for the forward engine; all three training seeds
  of the network); bars: weighted mean, numbers: RMS deviation (Table~\ref{tab:geant4}). (b, c) Arrival-time density for
  photons emitted next to the guide and at the center of one cell, all methods on the same bins (0.25~ns up to 2~ns, then
  1~ns and 2~ns).}
  \label{fig:validation}
\end{figure*}

Figure~\ref{fig:validation} and Table~\ref{tab:geant4} summarize the result. The adjoint agrees with \textsc{Geant4} in all
regions: the ratio is $0.999\pm0.003$, the RMS deviation of 1.9\% equals the \textsc{Geant4} statistical error, and
$\chi^2 = 37$ for 40 values. The forward engine agrees reasonably within its larger statistical error ($0.986\pm0.008$,
$\chi^2 = 30$ for 20 values). 

The network yields no offset ($1.002\pm0.004$), but its values scatter more, by 3.2\% RMS. Independent errors add in
quadrature, and this scatter splits into three parts, $3.2\% \approx \sqrt{1.9^2 + 1.8^2 + 1.9^2}\,\%$: the Geant4
counting noise (1.9\%), the variation between networks trained with different seeds (1.8\%), and an error common to all
three networks (1.9\%), which is the systematic error of the surrogate. Even with the seed variation counted as an
uncertainty, $\chi^2 = 68$ for 40 values; the excess is this common error.
Mean arrival times and arrival-time spreads agree to 2-3\% RMS for the adjoint and the network and
to 5-6\% for the forward engine, without offset, and the arrival-time distributions agree over three decades
(Fig.~\ref{fig:validation}b,c).
To compare the complete distributions $f_0(t)$, we normalize each over 0-60~ns, group it into 41 bins (0.25~ns wide up
to 2~ns, 1~ns up to 10~ns and 2~ns up to 60~ns), and compute the Jensen-Shannon divergence
\begin{equation}
  \mathrm{JS}(f,g) = \tfrac12\sum_k f_k\log_2\frac{f_k}{\mu_k} + \tfrac12\sum_k g_k\log_2\frac{g_k}{\mu_k},
  \label{eq:js}
\end{equation}
where $\mu_k = \tfrac12(f_k+g_k)$; JS is 0 for identical distributions and 1~bit for disjoint ones. \textsc{Geant4} detects only $10^3$ to $10^4$ photons per
region, so its counting noise alone makes JS a few $10^{-3}$~bits even for an exact model; we estimate this floor by
drawing samples of the actual \textsc{Geant4} size (and, for the forward engine, of its own size) from each method's
distribution. The adjoint's mean JS equals the floor ($2.6\times10^{-3}$~bits; Table~\ref{tab:geant4}), so its distributions cannot be
distinguished from \textsc{Geant4}'s at this statistics, and the forward engine is close to its higher floor (12.2 against
$11.3\times10^{-3}$). The network's JS value is 2.2 times the floor, and 63\% of it comes from the first 2~ns: its
distributions are built from 24 log-normal components, each about 34\% wide in time, which smooth the sub-nanosecond
structure of the prompt peak. The fraction of light arriving within 1~ns differs from \textsc{Geant4} by 1.5 percentage points RMS
per region, against 0.7 for the adjoint, which includes the \textsc{Geant4} noise. Measured by how far light is displaced in time
(the Wasserstein distance), this shape error is small: 0.15~ns for the network against a counting floor of 0.13~ns.
A finer test of the adjoint uses $2\times10^7$ \textsc{Geant4} photons emitted over the whole cell of two geometries and counted
per voxel and channel: $\chi^2/\mathrm{ndf} = 1.00$ and $1.03$ (6400 degrees of freedom each, binomial errors), and the
total captured fractions agree within 0.4\%.

\begin{table*}[t]
\centering
\small
\caption{Agreement with Geant4 for channel~0: captured fraction in 5 emission regions of 8 test geometries (40 values;
20 for the forward engine, run on 4 geometries). Ratio: mean of captured fraction of method/Geant4 weighted by the inverse variance, with its standard error; RMS: RMS of (method/Geant4 $-$ 1),
to be compared with the RMS Geant4 statistical error of 1.9\%; $\chi^2$: Eq.~(\ref{eq:chi2}). $\bar t$, $s_t$: RMS
relative difference of the mean arrival time and of the arrival-time spread; the mean differences are below 1\% for all
methods. JS: mean Jensen-Shannon divergence of the arrival-time distribution, Eq.~(\ref{eq:js}), with the value expected
from counting noise alone in parentheses. Network: mean over three training seeds.}
\label{tab:geant4}
\setlength{\tabcolsep}{7pt}
\begin{tabular}{lcccccc}
\hline
 & ratio & RMS (\%) & $\chi^2$ / values & $\bar t$ RMS (\%) & $s_t$ RMS (\%) & JS ($10^{-3}$ bits) \\
\hline
adjoint ($2\times10^5$ roots) & $0.999 \pm 0.003$ & 1.9 & 37 / 40 & 2.0 & 2.6 & 2.6 \;(2.6) \\
forward engine (JAX)          & $0.986 \pm 0.008$ & 5.3 & 30 / 20 & 4.7 & 5.7 & 12.2 \;(11.3) \\
network                       & $1.002 \pm 0.004$ & 3.2 & 68 / 40 & 3.1 & 2.8 & 5.8 \;(2.6) \\
\hline
\end{tabular}
\end{table*}

\textsc{Geant4} also tests the two approximations of the optical model. Tracking the polarization instead of averaging the Fresnel
reflectances changes the captured fraction by $-0.01\pm0.26\%$, i.e.\ not measurably. 

\subsection{Precision and cost of the adjoint field}
\label{sec:res_adjoint}
\begin{figure}[t]
  \centering
  \includegraphics[width=\columnwidth]{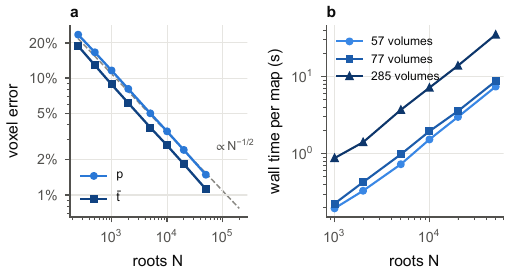}
  \caption{Adjoint field error and computing time versus number of roots $N$. (a) Voxel error of $p$ and $\bar t$ on the 80 test geometries.
  (b) Wall time per map on 14 cores for scenes of 57, 77, and 285 volumes.}
  \label{fig:adjoint}
\end{figure}

Tracing the 80 test geometries with $N = 250$ to $5\times10^4$ roots and independent seeds shows that the voxel error
falls approximately as $N^{-1/2}$, without bias (Fig.~\ref{fig:adjoint}a); at $2\times10^4$ roots it is 2.4\% for $p$
and 1.8\% for $\bar t$. Because the roots are quasi-random (a Halton sequence), the tracer's own error estimate, which
assumes independent roots, is too large, so reference errors are measured from independent replicate maps instead.
A $2\times10^4$-root map takes 3.0-3.6~s for typical scenes and 13.8~s for the most complex one
(Fig.~\ref{fig:adjoint}b). For these checks two settings matter. The first is a cap of 750 boundary interactions per root, which we used in earlier work; this cuts
long totally reflected paths and biases $\bar t$ by $-10\%$ and $s_t$ by $-19\%$; we use 5000 as the cap here. The second is Fresnel splitting, which is
unbiased but not efficient here: on four test geometries, pure analog reverse transport reaches nearly the same error (2.7\% instead of 2.5\%) in
a fifth of the time, because most interfaces send nearly all the weight one way.

\subsection{Surrogate accuracy}
\label{sec:res_surrogate}
Trained on 628 geometries, the network reaches a voxel error of 3.0\% for $p$, 1.9\% for $\bar t$ and 0.9\% for $s_t$
on the 80 test geometries (mean of three training seeds), comparable to a $2\times10^4$-root adjoint map (2.4\%, 1.8\%
and 1.2\%). The complete arrival-time distribution of each voxel is reproduced with a
Jensen-Shannon divergence, Eq.~(\ref{eq:js}), of $3.3\times10^{-3}$~bits ($p$-weighted mean over voxels), against
$2.1\times10^{-3}$ for a $2\times10^4$-root map, which is sampling noise; the reference's own noise contributes about
$0.2\times10^{-3}$. For the whole-cell distribution the network's divergence is $6\times10^{-5}$~bits, against
$1\times10^{-5}$ for the $2\times10^4$-root map: as for $p$, the network's errors are correlated between voxels and do
not average out.


The network's error differs in kind from the one of the adjoint. Part of it is an offset: the whole map of a test geometry is too high or too low by about the same factor, and this factor is large for only a few of the geometries. This offset is an interpolation error between designs: on training geometries the
networks reproduce the mean of each map to 0.3\%, closer than the 0.6\% statistical error of those maps, but on the
test geometries only to 2.3\% RMS, comparable to the common error of 1.9\% found against Geant4. 

Across the test set, neighboring-voxel errors have correlations of 0.6–0.8. Networks trained with different seeds also make largely the same errors (correlation 0.7), and averaging three networks reduces the error of p only from 3.0\% to 2.6\%. These correlated errors do not average out when the response is summed over many voxels: along random straight particle tracks, the network’s error is 2.5\%, against 1.4\% for a $2 \times 10^4$-root map.

\begin{figure*}[t]
  \centering
  \includegraphics[width=\textwidth]{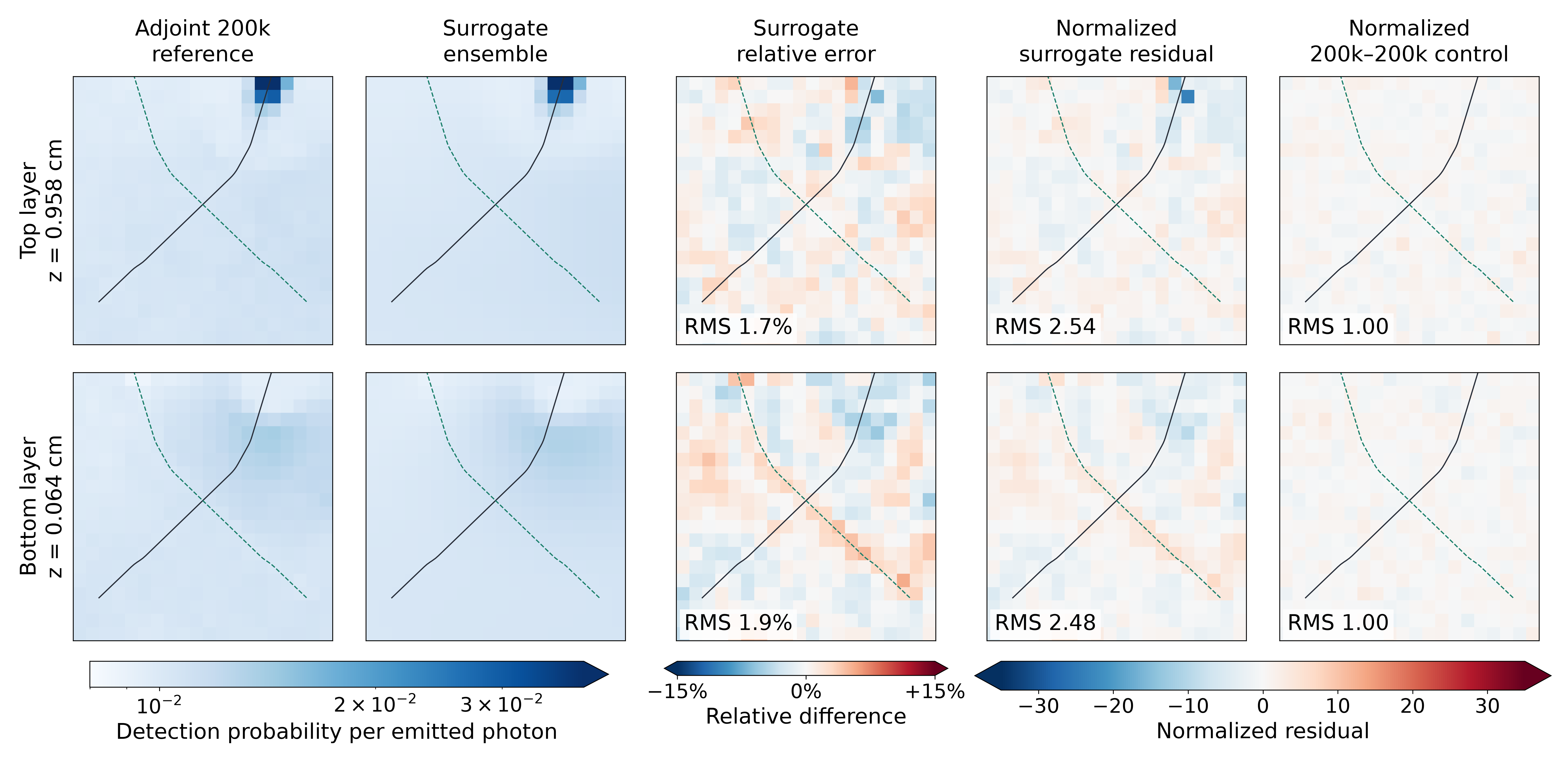}
  \caption{Detection probability on guide 0 for a test geometry, in the top and bottom layers: adjoint reference ($2 \times 10^5$ roots), three-network ensemble, relative difference to the reference, surrogate residual normalized to the reference noise, and normalized difference between two independent $2 \times 10^5$-root adjoint maps. Solid and dashed lines show the projections of guides 0 and 1, respectively.}
  \label{fig:maps}
\end{figure*}

Figure~\ref{fig:maps} compares the three-network ensemble
with the adjoint reference in a held-out geometry.
Let $S_v$ be the ensemble prediction and $A_v$, $B_v$
two independent adjoint estimates, each obtained with
$2\times10^5$ roots per geometry. For each depth layer $z$,
containing $N_z=400$ voxels, we define
\[
R(z)=\left[\frac{1}{N_z}
\sum_{v\in z}(A_v-B_v)^2\right]^{1/2},
\qquad
\sigma(z)=\frac{R(z)}{\sqrt{2}}.
\]
For independent replicas with equal variances, $\sigma(z)^2$
estimates the layer-average variance of a single reference
map. The normalized surrogate residual and replica control are
\[
P_v=\frac{S_v-A_v}{\sigma(z)},
\qquad
C_v=\frac{B_v-A_v}{R(z)}.
\]
The control has unit RMS within each layer by construction.
Since the normalization is shared by all voxels in a layer
and excludes surrogate uncertainty, $P_v$ measures discrepancies
relative to typical reference noise, rather than a voxel-specific
statistical significance.

The geometry shown illustrates good agreement rather than
median test-set performance. The observed $p$-weighted RMS
relative differences are 1.7\% in the top layer and 1.9\%
in the bottom layer, without subtraction of reference noise
in quadrature. The corresponding unweighted RMS values of
$P_v$ are 2.54 and 2.48. Thus, small relative discrepancies
still exceed the typical Monte Carlo noise of the reference.
Their spatial structure can be compared with the independent
adjoint-replica control, whose unit RMS follows from its
normalization and is not an independent uncertainty-calibration
test. This illustrative comparison uses a separate training
and evaluation campaign with the same sample counts as the
main study, but a newly generated split and three newly
trained networks.

\begin{figure}[t]
  \centering
  \includegraphics[width=\columnwidth]{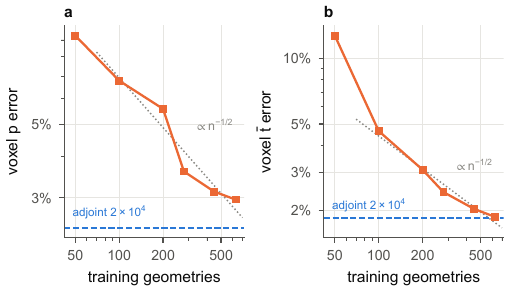}
  \caption{Network error versus number of training geometries for (a) detection probability and (b) mean arrival
  time, on the 80 test geometries; data points indicate the mean over three networks trained with different seeds on
  the same geometries. Dotted: $n^{-1/2}$ scaling, shown for comparison. Dashed: values relative to the adjoint map computed with $2\times10^4$ roots.}
  \label{fig:scaling}
\end{figure}

Pooled over the 80 test geometries, the error falls as $n^{-1/2}$ with the number of training geometries $n$,
for both $p$ and $\bar t$ (Fig.~\ref{fig:scaling}), and has not leveled off at 628, so a larger atlas will improve the
surrogate further. Replacing the recurrent guide encoder by a perceptron on the same vertex features gives the same
accuracy, so for guides with a fixed number of vertices the recurrent encoder is not essential.

\begin{figure*}[t]
  \centering
  \includegraphics[width=\textwidth]{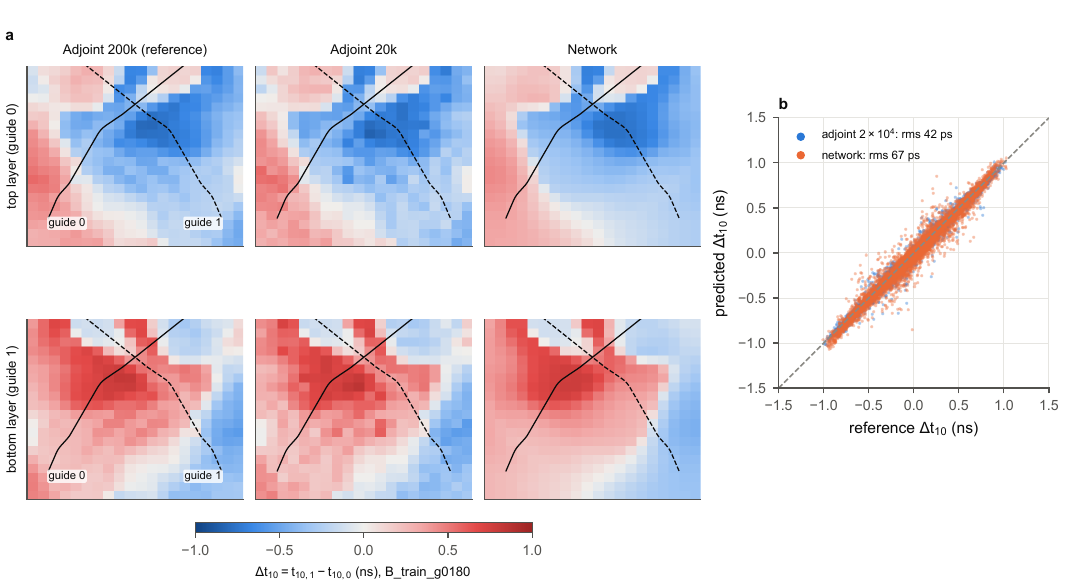}
  \caption{Difference of the early arrival times of the two readouts, $\Delta t_{10}$. (a) Top layer (next to guide~0) and bottom layer (next to guide~1)
  of one test geometry: reference ($2\times10^5$ roots), $2\times10^4$-root adjoint map and network. Solid line: guide~0,
  read out as channel~0; dashed: guide~1, channel~1; both readout ports lie beyond the upper edge of the maps.
  (b) Predicted versus reference $\Delta t_{10}$ for 8000 randomly chosen voxels of the 80 test geometries per method;
  legend: RMS difference over all voxels.}
  \label{fig:t10}
\end{figure*}

\begin{figure*}[t]
  \centering
  \includegraphics[width=\textwidth]{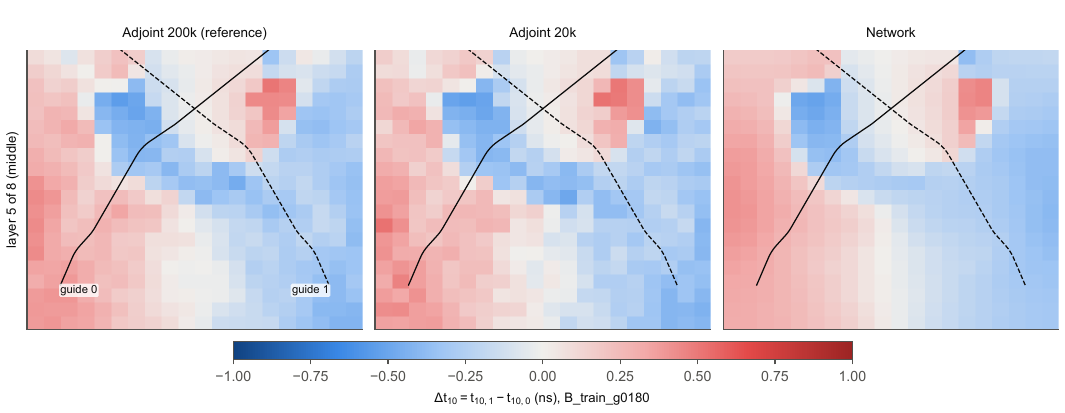}
  \caption{$\Delta t_{10}$ in layer 5 of 8, just above mid-depth, of the geometry of Fig.~\ref{fig:t10}; columns and
  guides as in Fig.~\ref{fig:t10}a.}
  \label{fig:t10_mid}
\end{figure*}

\begin{figure}[t]
  \centering
  \includegraphics[width=\columnwidth]{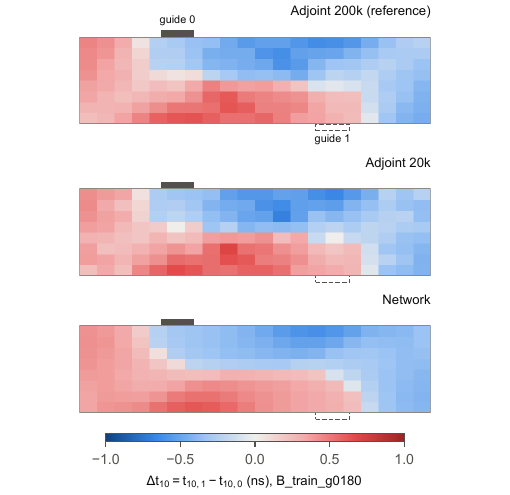}
  \caption{$\Delta t_{10}$ in a vertical ($x$-$z$) section at $y = 2.23$~cm (voxel row 11 of 20) of the geometry of
  Fig.~\ref{fig:t10}: reference, $2\times10^4$-root map and network, top to bottom. Filled bar: guide~0 on the top face;
  dashed outline: guide~1 on the bottom face, where they cross the section.}
  \label{fig:t10_xsec}
\end{figure}

The surrogate also reproduces the difference between the early arrival times of the two readouts,
$\Delta t_{10} = t_{10,1} - t_{10,0}$, where $t_{10,c}$ is the time by which readout $c$ has received 10\% of its light.
Because the two guides are mirror images, the response of readout~1 at a point equals that of readout~0 at the mirrored
point ($x \to X - x$, $z \to Z - z$); $\Delta t_{10}$ therefore changes sign under this mirroring and vanishes on its
axis. Away from the axis the two readouts see an emission point from different distances and directions, and
$\Delta t_{10}$ reaches $\pm0.8$~ns (RMS 0.36~ns over the test geometries), which a two-ended readout can use to locate
a deposit. Figure~\ref{fig:t10}a shows it in the top and bottom layers of one test geometry, Fig.~\ref{fig:t10_mid} in a
layer near mid-depth and Fig.~\ref{fig:t10_xsec} in a vertical section. Near the side walls it is set by the horizontal
position, while in the interior each readout receives its earliest light from the half of the cell away from its own
guide: the top layer, next to guide~0, is mostly negative and the bottom layer mostly positive. \textsc{Geant4}, which simulates
both readouts independently, shows the same dependence on depth: for uniform emission in each layer of this geometry,
$t_{10}$ of readout~0 rises from 0.38~ns in the bottom layer to 0.43~ns in the top layer, against 0.39 and 0.43~ns for
the adjoint. The network reproduces the regions of either sign and their boundaries in all three views, but smooths
structures a few voxels wide that the reference is instead able to resolve; the $2\times10^4$-root map keeps that structure with added voxel
noise. On the 80 test geometries the RMS error of $\Delta t_{10}$ is 67~ps for the network and 42~ps for a
$2\times10^4$-root map, with no visible bias for either (Fig.~\ref{fig:t10}b).

Ten further designs are used to test a guide shape that is absent from the training and testing sets: straight guides without bends. Their cell
dimensions, guide widths and fillet ratios span the design ranges, and in parameter space they are no farther from the
training designs than the test geometries are, but every training guide bends, by at least $17^\circ$ at one vertex.
Because start and end points are fixed fractions of the cell, a straight guide always runs along the same diagonal, and
the two mirrored guides cross as an X in top view. On these designs the network reproduces the timing as well as on the
test geometries ($\bar t$ 2.0\%, $s_t$ 1.0\%, $\Delta t_{10}$ 70~ps), and its error of $p$ is 3.5\%. The network thus carries the
arrival-time structure over to a guide shape it has not seen within a few percent accuracy.


\subsection{Geometry derivatives}
\label{sec:res_derivatives}

The fact that our surrogate can be differentiated does not in general guarantee that its geometry derivatives match those of the transport. In six
test geometries we varied five design parameters (cell thickness, guide width, bend radius, and the $x$ and $y$ positions of a
turning point) along a line, $u = u_0 \pm 0.04$, $\pm 0.08$ and $\pm 0.16$ in units of the design range wherever the
design stays valid, rebuilt the geometry at each point, and traced it with eight independent seeds of
$5\times10^4$ roots, sharing random numbers between the points. The adjoint derivative is the slope of a least-squares
line through the symmetric points or, where the design range or the fillets exclude one side, of a quadratic through the
valid points; for six derivatives near such a limit only a central difference with a step of 0.015 or 0.04 is available.
A five-point stencil agrees with the central differences within their statistical error, so the truncation error is
negligible; the remaining uncertainty is statistical, 0.017 RMS for $\partial\ln\bar p/\partial u$ and 0.09~ns for
$\partial\bar t/\partial u$. From the whole-cell arrival-time distribution (uniform emission, transport only) we take as diagnostic metrics the
mean $\bar t$ of the arrival distribution, its standard deviation $s_t$, and the 10\% quantile $t_{10}$, and differentiate the network in the same way.
We report the agreement separately for the two parameter families: the main dimensions (cell thickness and guide width)
and the guide path (bend radius and turning point).

\begin{figure}[t]
  \centering
  \includegraphics[width=\columnwidth]{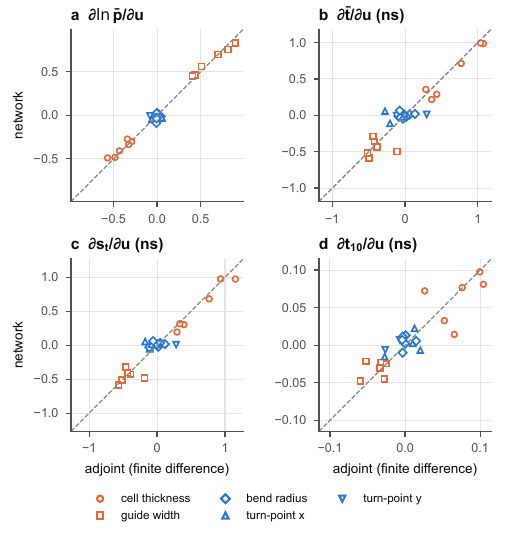}
  \caption{Derivatives with respect to five design parameters in six test geometries, network versus adjoint finite
  differences: (a) logarithm of the whole-cell detection probability $\bar p$; (b) mean $\bar t$, (c) standard deviation
  $s_t$ and (d) 10\% quantile $t_{10}$ of the whole-cell arrival-time distribution (uniform emission over the cell,
  channel~0, 0-60~ns). $u$: parameter in units of its design range. Orange: cell thickness and guide width; blue: guide
  path. Dashed line indicates equality of the two estimates.}
  \label{fig:derivatives}
\end{figure}

For the logarithm of the collected light (Fig.~\ref{fig:derivatives}a) the correlation is 0.99, with an RMS difference of
0.05 against an RMS derivative of 0.39. For the main dimensions the derivatives agree in sign and to 9\% RMS. The small
sensitivities to the guide path (up to 0.08 per unit of the design range) are not reproduced: the correlation within this
family is 0.1, and the network has the wrong sign for four of twelve.

Derivatives of the mean arrival time agree less well (correlation 0.94, RMS difference 0.15~ns against an RMS derivative
of 0.44~ns), for two reasons. First, the reference is noisier: over the scanned range the mean arrival time changes by
only a few hundredths of a nanosecond, and the statistical error of the reference accounts for about 0.09~ns of the
difference. Second, the geometry changes the arrival time less, relative to the network's error, than it changes the
collected light: across the test geometries the whole-cell mean arrival time varies by 0.33~ns (s.d.) against a network
error of 0.04~ns, a ratio of 9, compared with 15 for $\ln p$. A derivative inherits the error of the function divided by
the distance over which that error varies, so the timing derivatives suffer deviations first. For the main dimensions, which move
the mean arrival time by up to 1.1~ns per unit of the design range, the network estimate has the right sign in all twelve cases
and a median deviation of 17\% (Fig.~\ref{fig:derivatives}b). The guide-path sensitivities of the reference are 0.15~ns
per unit RMS and up to 0.30~ns; the network predicts 0.05~ns RMS, with a correlation of 0.15 and in some cases the
opposite sign. The guide path affects the timing more than the yield, because the light travels along it.

The width of the distribution follows the mean: in the adjoint, $\partial s_t/\partial u$ and $\partial\bar t/\partial u$
have a correlation of 0.99 and a slope of 1.0, because the design changes mainly stretch the late tail, and the network
reproduces this (Fig.~\ref{fig:derivatives}c); its difference from the reference, 0.12~ns RMS, is comparable to the
reference's statistical error. The early edge hardly moves: the whole-cell $t_{10}$ is 0.3-0.45~ns, and its derivatives
are about ten times smaller than those of $\bar t$ (0.04~ns per unit RMS). The network has the right sign for all twelve
derivatives with respect to the main dimensions, with a median deviation of 26\%, but not for the guide path
(correlation 0.89 overall; Fig.~\ref{fig:derivatives}d).

The network is seen to correctly account for the guide: across the 80 test geometries it reproduces how the mean arrival time grows with
guide length (0.062-0.067~ns/cm, against 0.062~ns/cm for the adjoint). The network thus learns the trend between designs; however it struggles with the local slope at a given design. Its gradients can therefore steer the main design dimensions of the yield and, less
precisely, of the timing; changes of the guide path and timing optimization should use the adjoint, with steps of several
percent of the design range and more roots for timing derivatives than for yield derivatives. Overall, the discussed performance suggests that the surrogate provides a useful approximation of the system in the way of a full optimization by gradient descent; however, a confirmation will require its actual use in an optimization pipeline, which we defer to future work.

\subsection{Cost of simulating events and maps}
\label{sec:res_cost}

The relevant task for a detector simulation is, given the energy deposits of an event, to produce the arrival-time
distribution at each readout. Forward transport follows every photon from generation to extinction, loss, or collection, so its cost grows with the number of photons $N_\gamma$. With a response field the task is summarized by Eq.~(\ref{eq:superposition}): a weighted sum of the fields at the deposit
positions, convolved with the emission profile of EJ-204 (0.7~ns rise, 1.8~ns decay), followed by one multinomial draw of the detected photons per time bin.
Its cost depends on the number of deposits rather than on $N_\gamma$, but the field must first be generated. Which method is
cheaper therefore depends on which costs are counted in the comparison. Here we distinguish three components. The one-time cost is that of building the network: producing an atlas of 628 adjoint maps (37~min) and running the training on those data (23~min) add up to 1.0~h of computing time; this is shared by all geometries of the design
family. The per-geometry cost is the one of computing one adjoint map, 3.6~s for $2\times10^4$ roots; the network does not incur in that cost. The per-event
cost, for a single charged particle crossing the cell (which for a minimum ionizing particle corresponds to approximately 26,500 photons), is 0.4~ms with an adjoint map and 2~ms with the
network evaluated at the deposits. \textsc{Geant4} with the same optical model and a compiled photon loop tracks
$1.7\times10^5$ photons per second on the 14 cores, hence spends 0.16~s for such an event.

\begin{figure*}[t]
  \centering
  \includegraphics[width=\textwidth]{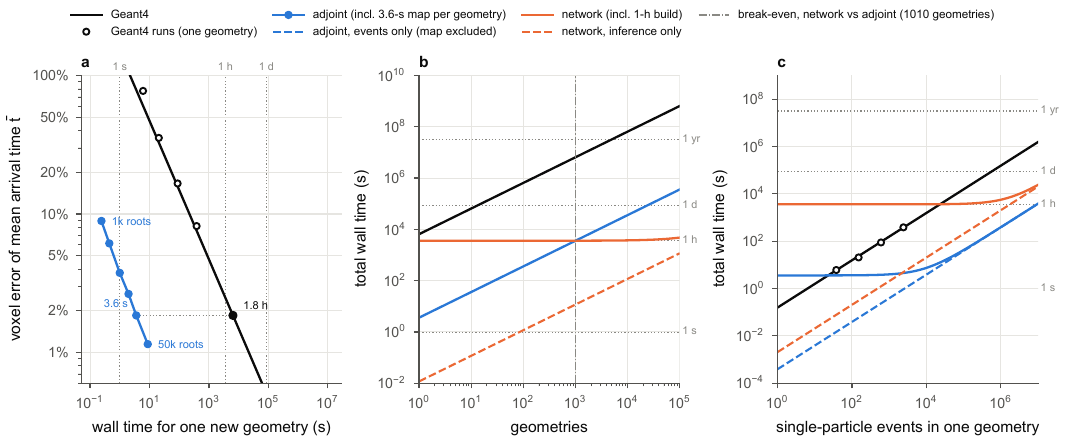}
    \caption{Cost of the methods on matched tasks, on the same 14-core CPU. Solid lines include the one-time costs (the
  adjoint map of each geometry, the network's 1h build of atlas and training); dashed lines leave them out.
  (a) Wall time for one new geometry versus the voxel error of the mean arrival time $\bar t$ ($p$-weighted RMS relative
  difference to the $2\times10^5$-root reference). \textsc{Geant4}: uniform emission, each detected photon assigned to the voxel
  where it was emitted; line: error from the number of detected photons per voxel at the measured photon rate (median of
  the test geometries); open circles: \textsc{Geant4} runs for one geometry. Adjoint: measured runs of 1000 to $5\times10^4$ roots. Filled circle: \textsc{Geant4} at the accuracy of a $2\times10^4$-root map.
  (b) Estimated total time for multiple geometries at that accuracy; dash-dotted: break-even of network and adjoint.
  (c) Total time for multiple single-particle events (26,500 photons) in one geometry; open circles: the \textsc{Geant4} runs of (a),
  placed at their photon number divided by 26,500. The \textsc{Geant4} energy deposition, common to all, is not included.}
  \label{fig:cost_matched}
\end{figure*}

Figure~\ref{fig:cost_matched} compares the methods at the same accuracy, measured by the voxel error of the mean arrival
time, which is harder to obtain by forward simulation than the detection probability. For one new geometry
(Fig.~\ref{fig:cost_matched}a), a forward map as accurate as a $2\times10^4$-root adjoint map or the network (about 1.9\% in $\bar t$) needs
about $1.1\times10^9$ emitted photons: 1.8~h with \textsc{Geant4} on 14 cores and 12 days with the forward engine on 3 cores,
against 3.6~s for the adjoint; the network reaches the same accuracy in 12~ms once it is built. For a scan of
geometries (Fig.~\ref{fig:cost_matched}b), the network's one-hour build pays off against the adjoint only beyond about
1000 geometries. For events in one geometry (Fig.~\ref{fig:cost_matched}c), the adjoint map is cheaper than \textsc{Geant4}
after just 23 events. Without counting its build, the network is cheaper than the adjoint for fewer than about 2200 events per
geometry, because it needs no map, and more expensive beyond, because each event costs 2~ms instead of 0.4~ms; counting
its build, the network is cheaper than \textsc{Geant4} only beyond about 23,000 events. The surrogate is therefore not always the faster
option: it pays off when many geometries are evaluated with few events each, as in a design scan or an optimization,
whereas for many events in a fixed geometry an adjoint map is cheaper. On a GPU the forward engine is faster than on a
CPU (Fig.~\ref{fig:g4_vs_jax}), but its cost still grows with $N_\gamma$; we have not run a matched GPU benchmark of all methods on one machine.

The per-event cost grows with the number of deposits rather than of photons. With the adjoint map it rises from 0.4~ms
for one particle (11 active voxels) to 1.4~ms for 1000 particles crossing the cell together ($2.7\times10^7$ photons,
3036 active voxels), since deposits are first summed per voxel and at most 3200 voxels can be active. The network
evaluated at every deposit grows from 2~ms to 0.29~s over the same range; evaluated once on the voxel grid (18~ms) and
combined in the same way, its cost no longer depends on the event.

\begin{figure*}[t]
  \centering
  \includegraphics[width=\textwidth]{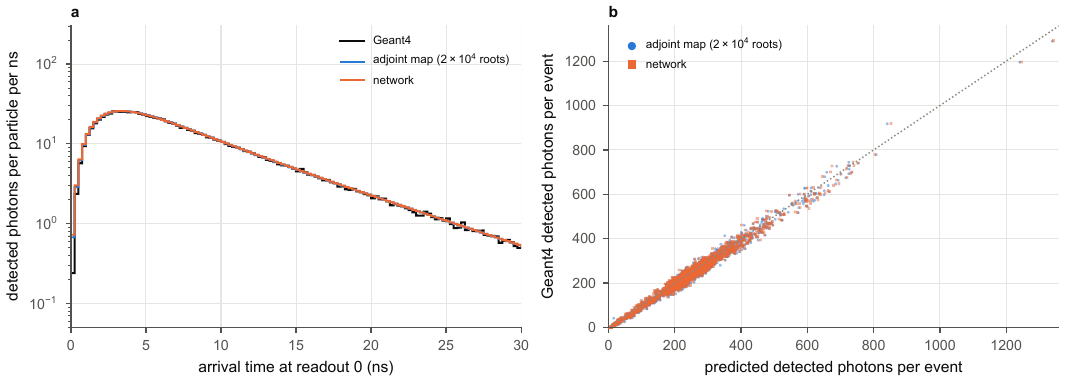}
  \caption{Events simulated end to end: 1000 minimum ionizing particles cross a geometry not used in training;
  \textsc{Geant4} gives the energy deposits ($2.65\times10^7$ scintillation photons in total) and the light is
  transported by \textsc{Geant4}, the $2\times10^4$-root adjoint map and the network. (a) Arrival-time distribution at readout~0,
  summed over all events, on 0.25-ns bins. (b) Detected photons per event and channel, \textsc{Geant4} versus the prediction
  $\lambda$; the diagonal indicates equality of the estimates.}
  \label{fig:events}
\end{figure*}

Figure~\ref{fig:events} shows an end-to-end test. A thousand charged particles in minimum ionizing regime 
cross a scintillator with a geometry not used in training;
\textsc{Geant4} provides the energy deposits ($2.65\times10^7$ scintillation photons), and the light is
transported by \textsc{Geant4} and predicted by the adjoint maps and the network. The arrival-time distributions summed over all
events agree over three decades (Fig.~\ref{fig:events}a). In total \textsc{Geant4} detects 244,107 photons in channel~0; the
$2\times10^5$-root adjoint map predicts 0.15\% more, the $2\times10^4$-root map 1.5\% more, and the network 1.6\% more,
each with a statistical error of 0.2\%. Event by event (Fig.~\ref{fig:events}b), the detected counts scatter around the
prediction as expected from photon statistics
, and the mean arrival time of
an event differs from \textsc{Geant4} by 0.54~ns RMS, {\em i.e.} no more than its own statistical fluctuation (0.53~ns). In this geometry the
per-event differences are therefore set by photon statistics; the network's 1.6\% excess is within its per-geometry
error (2.3\% RMS, Sec.~\ref{sec:res_surrogate}).

The two approaches answer slightly different questions. A forward simulation produces one random realization of the event, including the correlations between photons from the same deposit. The response field gives the expected
intensity, with a statistical precision set by the map (2.4\% per voxel, 0.6\% for the whole cell at $2\times10^4$
roots); detector fluctuations are added by sampling from it; the sampling is included in the timing estimates discussed above.

\section{Discussion}
\label{sec:conclusions}

Tracing light backward from the readout face of a light guide coupled to a scintillator block turns the expensive part of optical simulation --finding the rare paths that
reach a small port-- into a much cheaper procedure. For the cell studied here, one adjoint run of $2\times10^4$ roots gives the detection
probability and arrival-time distribution of every emission point with 2.4\% accuracy using 3.6~s of a 14-core CPU, and agrees with \textsc{Geant4} within
the statistical precision of the comparison, both in the collected light and in its timing. Once such a field exists, simulation of the
readout signal of an event costs the same computing time for $10^3$ or $10^{10}$ photons.

The optical model entails one approximation that matters at the few-percent level: photons that leave into air are terminated;
allowing them to re-enter raises the collected light by 3.5\%. Polarization has instead no measurable effect. Two practical points concern specifically the adjoint method. The first is that a step cap that is too short biases the timing distribution far more than it affects the overall yield. The second is that the analog reverse
transport is more efficient than Fresnel splitting in this geometry.

The surrogate predicts the field of a new design in about 10~ms, to 3.0\% per voxel for the detection probability
(2.6\% for an average of three networks) and 1.9\% for the mean arrival time on 80 test geometries. It reproduces the full
arrival-time distributions well, with minor discrepancies observed in the sub-nanosecond structure of the prompt peak. The surrogate also correctly models the early-arrival
asymmetry between the channels to 67~ps; its main source of error is systematic rather than statistics-driven: this is partly traced back to a per-geometry offset, which
anyway decreases with the size of the training atlas. Computed gradients follow the transport for the main design dimensions, cell
thickness and guide width (to 9\% RMS for the collected light, with a median deviation of 17\% for the mean arrival
time); these metrics do not show sensitivity to small details of the guide path, for either light or timing. After its one-hour build on the considered training geometries, the computed surrogate pays off over the adjoint model when many geometries are evaluated with few events each, as is typical in an optimization design scan; for many events in one geometry the adjoint map remains instead cheaper. The surrogate is therefore suited to fast exploration and coarse optimization of the main dimensions, while changes of the guide path, timing optimization and final designs should use the adjoint, which at a few seconds per geometry fits inside an optimization loop.

Future extensions of this work include the detailed modeling of photon re-entry through air in the fast models, a geometry-level term in the surrogate model to remove its offset, larger atlases for a more precise training of the surrogate model, a wider consideration of different guide topologies and surface models, and a GPU implementation of the adjoint tracer. 

\begin{acknowledgments}

PV is supported by the “Ramón y Cajal” program under Project No. RYC2021-033305-I funded by the MCIN MCIN/AEI/10.13039/501100011033 and by the European Union NextGenerationEU/PRTR. All authors except LP are supported by the European Innovation Council (EIC) Pathfinder project PHINDER, grant agreement No. 101258353, funded by the European Union.

The authors gratefully acknowledge the computer resources at Artemisa and the technical support provided by the Instituto de Física Corpuscular, IFIC (CSIC-UV). Artemisa is co-funded by the European Union through the 2014-2020 ERDF Operative Programme of Comunitat Valenciana, project IDIFEDER/2018/048.

This work has benefited from activities carried out within the framework of the COST Action MLQC4FC (CA24146), supported by COST (European Cooperation in Science and Technology).

\begin{figure}[h!]
\includegraphics[width=0.9\linewidth]{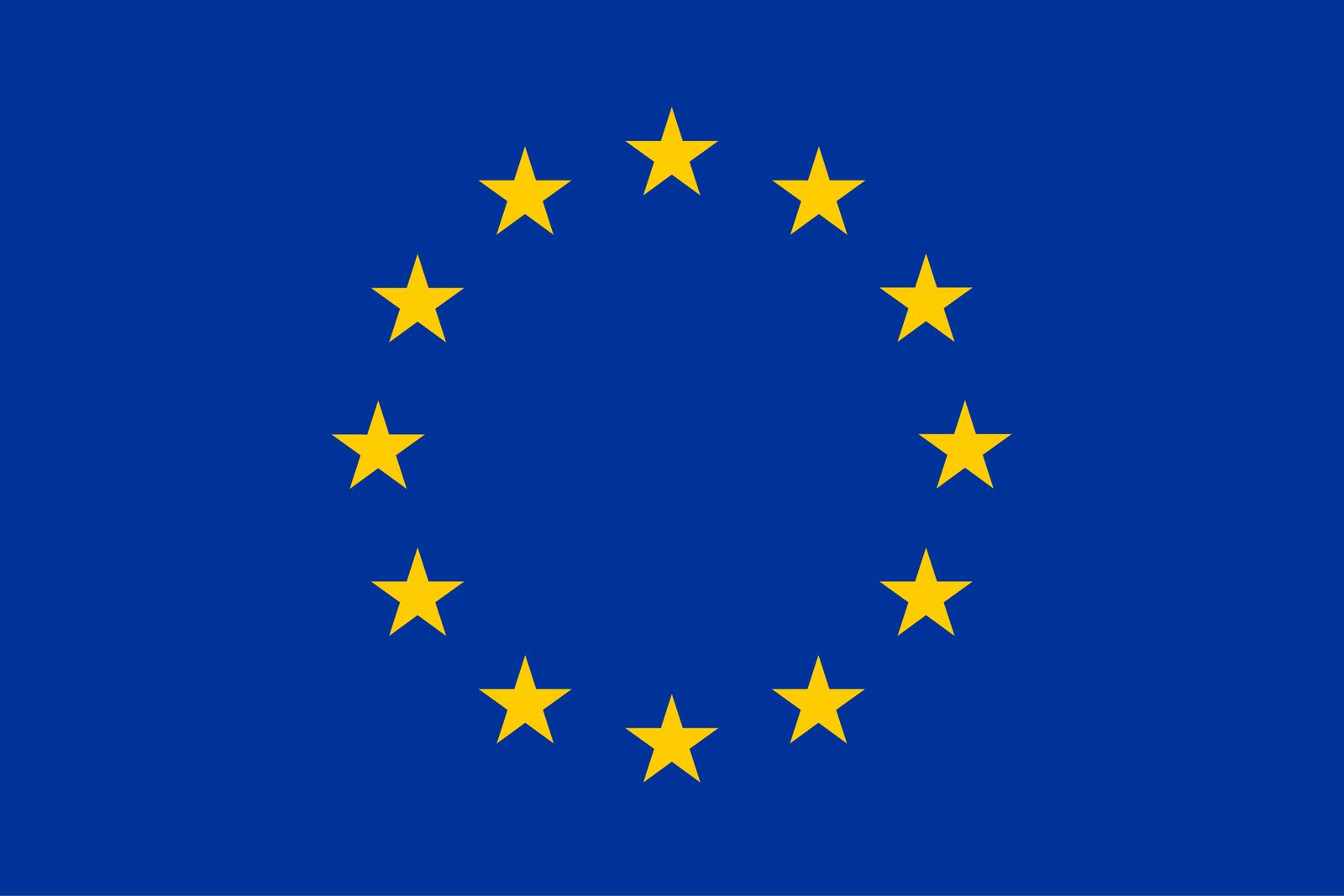}
\end{figure}

\end{acknowledgments}

\section*{Data and code availability}

The data are archived on Zenodo under the Creative Commons Attribution 4.0 licence
(\href{https://doi.org/10.5281/zenodo.24459410}{doi:10.5281/zenodo.24459410}). 

The code is available in the
\href{https://github.com/PHINDER-EIC/adjoint-and-neural-surrogat-for-scintillator-detector}
{PHINDER GitHub repository}
under the MIT licence.
It contains
the scripts that generate the data, train the
networks and run the Geant4 validation, the analyses behind the reported numbers, and the scripts that produce the data figures and
Table~\ref{tab:geant4} from the archived data.

\appendix

\section{\textsc{Geant4} configuration}
\label{app:geant4}

\textsc{Geant4}~11.4.1 was used through geant4\_pybind~0.1.3~\cite{geant4pybind}. Each geometry was exported from the geometry builder
used by the forward engine and the adjoint tracer as closed triangle meshes (G4TessellatedSolid): the scintillator box; for
each guide the union of its core segments and terminal; and a cladding shell made of the outer cladding surface, the core
side faces with reversed orientation and two end rings. Core and cladding are touching volumes in an air world, so the core
end faces meet air as in the original scene. The physics list contains only G4OpticalPhysics with Cherenkov, scintillation,
wavelength shifting, Rayleigh and Mie processes disabled; boundaries carry no optical surface, so G4OpBoundaryProcess applies
the Fresnel and Snell laws for polished dielectrics with polarization. Photons have 3.04~eV, isotropic directions and random
linear polarization. A photon is detected when a step inside a core ends on that guide's readout plane. In the matched
configuration a photon that makes a step of non-zero length in air is killed (a reflected photon makes a zero-length step
whose pre-step volume is the air and is kept), and the polarization is redrawn perpendicular to the direction after every
boundary step. In the polarized configuration the latter is omitted, and in the re-entry configuration photons in air are
also kept.
For the events of Sec.~\ref{sec:res_cost} the same scene was built with real material compositions (polyvinyltoluene,
polystyrene core, PMMA cladding, G4\_AIR) and the standard electromagnetic physics constructor (G4EmStandardPhysics, data set
G4EMLOW~8.8) with 0.7-mm production cuts; the particle definitions were created in the physics list itself, because the
particle set of the standard constructors fails to initialize in this build. Steps with energy loss inside the scintillator
were written out (pre- and post-step position and time, deposited energy, step length, particle), and the scintillation
photons generated from them were tracked by a separate optical run of the matched configuration above.
\section{Run time of the forward engine on GPU}
\label{app:figures}

\begin{figure}[h]
  \centering
  \includegraphics[width=\columnwidth]{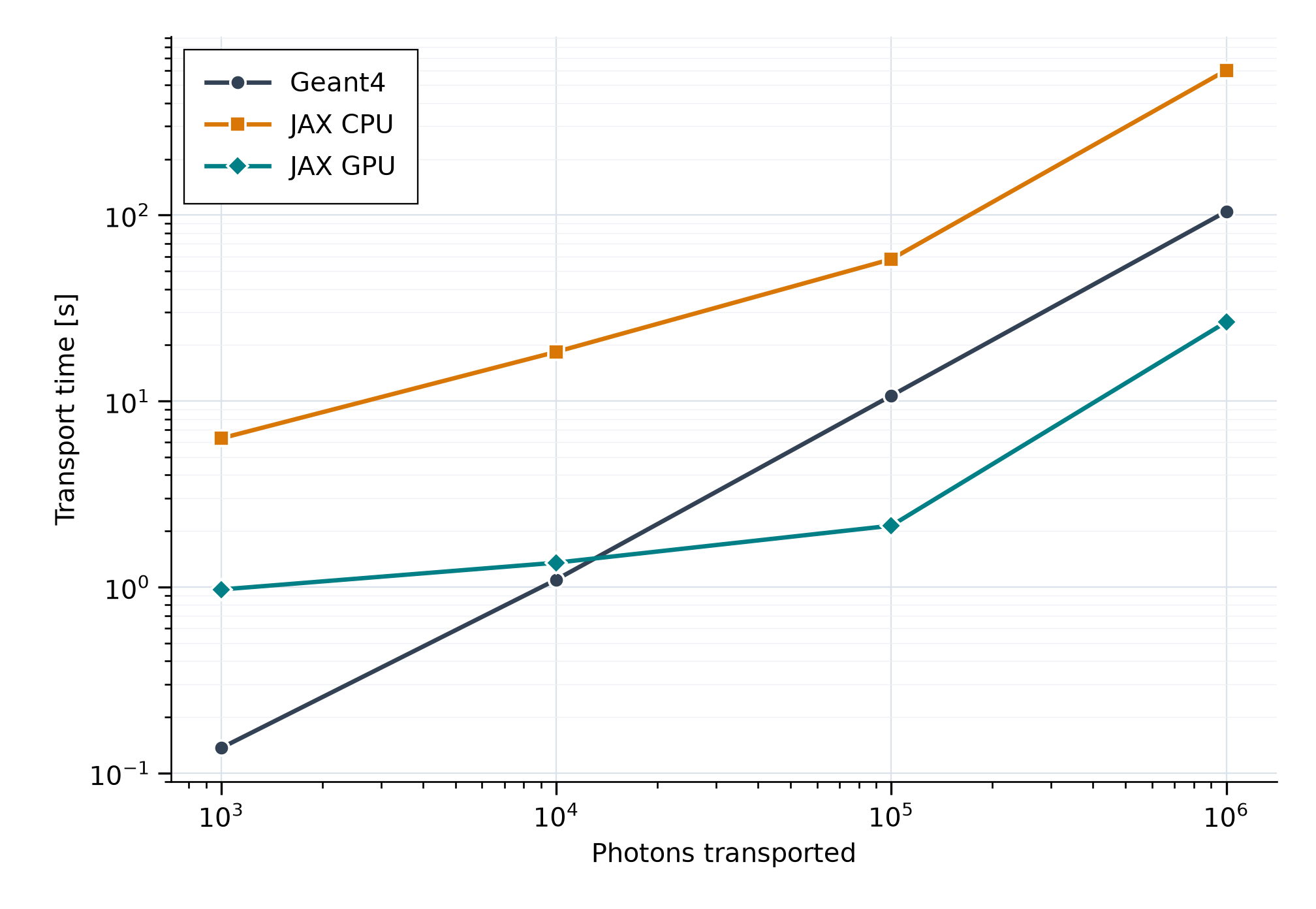}
  \caption{Transport time of \textsc{Geant4} and of the forward engine (JAX) on CPU and GPU for $10^3$-$10^6$ photons emitted from a
  point in the scintillator, on a workstation with an AMD Ryzen~9950X CPU and an NVIDIA GeForce RTX~5090 GPU (not the
  machine used elsewhere in this paper). Times include the transfer of results to the host and exclude compilation and
  warm-up.}
  \label{fig:g4_vs_jax}
\end{figure}

Figure~\ref{fig:g4_vs_jax} compares the run time of the forward engine with that of \textsc{Geant4} for photons emitted from one
point. For small runs both JAX versions are dominated by a fixed overhead, about 1~s on the GPU and 6~s on the CPU. For
large runs all three times grow roughly in proportion to the number of photons: on the CPU the forward engine is about
six times slower than \textsc{Geant4}, whereas on the GPU it is faster than \textsc{Geant4} above about $10^4$ photons, by a factor of four
to five at $10^5$-$10^6$ photons. These timings were made on a different machine from all other results and are not
used in the cost comparison of Sec.~\ref{sec:res_cost}.

\bibliography{references}

\end{document}